# Children's Online Safety Risks and Ethical Considerations in XR Games

Zinan Zhang
Information Sciences and Technology, The Pennsylvania State University,
University Park, PA, USA
zzinan@psu.edu

Xinning Gui
Information Sciences and Technology, The Pennsylvania State University,
University Park, PA, USA
xinninggui@psu.edu

Yubo Kou
Information Sciences and Technology, The Pennsylvania State University,
University Park, PA, USA
yubokou@psu.edu

## ABSTRACT

Emerging extended reality (XR) technologies are reshaping how children play, learn, and socialize—yet they also present serious safety risks. Gaming, a primary form of entertainment for children, is also one of the key applications of XR. While XR platforms offer immersive and engaging gaming experiences, recent news has highlighted safety concerns such as car accidents, lower judgment for real-world situations, and exposure to disturbing content like virtual rape. This research examines how XR game design may lead to online safety risks for children. Through analysis of player forums, game developer forums, and interviews with child players, we identify harmful XR design patterns, explore how developers collaboratively generate and implement risky game ideas, and document children's firsthand experiences of online safety risks. Existing ethical frameworks often fail to address the immersive and socially dynamic nature of XR games. We advocate for a child-centered, design-aware approach to ethical considerations in XR games, urging platforms and policymakers to prioritize children's developmental needs. Our work aims to help shape safer, more inclusive XR environments through research and cross-sector collaboration.

## POSITION STATEMENT

Emerging technologies such as extended reality (XR) are rapidly transforming how children engage with game play [9], health [4], education [17], and social interaction [1]. XR refers to all real-and-virtual combined environments and human-machine interactions generated by computer technology and wearables (e.g. Virtual Reality, Augmented Reality, and Mixed Reality). These technologies bring joy and expand the boundaries of learning and development.

One prominent application of XR is in gaming. XR platforms now host millions of games that children can explore for entertainment. From mobile AR games like Pokémon Go, which was downloaded over 500 million times in its first year [20], to fully immersive VR systems such as PlayStation VR and Move controllers [21], the landscape of extended reality XR games has expanded rapidly. Major tech companies like Logitech have also joined the ecosystem, offering XR-enhancing tools such as racing wheels and driving force shifts [22]. Game platforms like Rec Room [23] and Roblox [24] are increasingly adapting their ecosystems to XR, opening new forms of social and gameplay experiences.

However, alongside these innovations, XR technologies raise serious ethical concerns for children's well-being. Popular XR games illustrate some of these risks. Pokémon Go, for instance, has been linked to car accidents and pedestrian injuries due to distracted gameplay [25]. Simulated driving experiences may foster a false sense of security, reducing players' ability to react in real-world scenarios that demand quick judgment [8]. More disturbingly, immersive XR environments have enabled virtual reenactments of real-world harms, including instances of virtual rape [14], Nazi roleplay [12], and reenactments of mass shootings [5].

Despite these emerging risks, research on the risks associated with XR-based play remains limited. Scholars have begun to examine the broader implications of XR on children's development, including the erosion of physical community bonds, long-term impacts on health and development, and risks to children's safety and privacy [11]. Parents, educators, and healthcare professionals also express growing unease about how to manage content appropriateness, balance screen time, and monitor behavioral changes resulting from extended XR use [10]. While existing work has examined general concerns with XR technologies—including some within gaming contexts [6, 10, 11]—few studies have systematically explored the specific risks that children face within XR gameplay. As games continue to be a dominant and naturally appealing form of entertainment for children, this oversight represents a critical research gap.

As XR technologies become increasingly embedded in children's everyday play, there is a growing need to examine not just how these immersive platforms are used—but how they are designed. Design decisions shape what players can do, how they interact, and what kinds of experiences they have—including harmful ones. XR games offer powerful and engaging experiences, but when designed without adequate guardrails, they can expose children—one of the most vulnerable populations—to serious risks. Among the many XR platforms available today, Roblox stands out as both one of the most influential and widely adopted by children—and one of the most controversial.

While often perceived as a traditional online platform, Roblox [26] has rapidly expanded into the XR space. It now supports virtual reality gameplay and is part of a broader XR ecosystem where children can explore, build, and interact in fully immersive 3D environments through computers, mobile devices, or VR headsets. As of 2025, there are nearly 90 million daily active players on Roblox [27]. Approximately 39% of Roblox players are under the age of 13 [27], and 58% are under 16 [2]. Just a week after launching on Meta's Quest VR headset, Roblox was downloaded over 1 million times [15]—signaling its growing presence in the XR ecosystem. These numbers highlight the urgency of understanding how Roblox, as an emerging XR platform, shapes child players' experiences through its game design. Unlike conventional games built by professional studios, most Roblox games are developed by individual game developers. Millions of game developers develop countless XR games on Roblox. This decentralized model has led to an explosion of creativity, but also makes it extremely difficult for Roblox's moderation team to effectively address the hundreds of child safety issues that arise daily across millions of games [3, 16].

In response to the growing safety concerns on Roblox, particularly within its XR game design, our research investigates the children's safety risks embedded in Roblox game designs. We focus on how the design of XR games can lead to harmful experiences for children. Through our ongoing work, we have identified four recurring harmful design patterns: ubiquitous microtransaction design, unconstrained social design, unmoderated expression design, and problematic world design [13].

To better understand how harmful design patterns emerge, we conducted an analysis of Roblox game developers' online forums. Our findings revealed that some developers not only share risky or questionable design ideas but also exchange strategies for implementing them [19]. This discovery raises critical ethical concerns regarding the creation of child-friendly XR games. Specifically, we are prompted to ask: How should ethics be defined in XR games intended for children? And how can game developers and game companies be encouraged—or even required—to recognize and adhere to these ethical considerations?

While scholars and child advocacy organizations have proposed various ethical frameworks to guide digital innovation for young users—such as UNICEF's Responsible Innovation in Technology for Children [28], and the Age-Appropriate Design Code [7]—these initiatives often focus on broader technology ecosystems or web platforms. Recent scholars have also begun to address ethical concerns specific to XR applications, calling for greater attention to contextual considerations for emerging technologies [10]. However, there is currently no comprehensive ethical framework tailored specifically to XR games designed for children's play.

To further understand the primary players and contribute to the ethical framework of XR games, we examined how children engage with and experience safety risks in these XR games. Through an interview study with child players who play XR games like Roblox and RecRoom, our findings pointed to several safety risks, including exposure to scams, normalization of antisocial behavior, and interactions with adult roleplay that may not be age-appropriate [18]. These conversations offered powerful insights into the ethical considerations in creating child-friendly XR games and underscored the urgent need for platforms and policymakers to prioritize children's values and developmental needs when shaping policy and safety regulations.

I am eager to participate in this workshop because of my ongoing research on children's safety in XR—particularly in XR-based games. As immersive technologies become more deeply embedded in children's everyday play, there is an urgent need to critically examine how design decisions within XR games shape their experiences, including risks and harms. Some XR games like Roblox present especially complex ethical challenges due to their decentralized game development and lack of consistent oversight. While existing digital ethics frameworks offer important guidance, they often fall short when applied to the distinct, immersive, and socially dynamic environments of XR games for children. This workshop presents a timely opportunity to engage with interdisciplinary scholars who are equally committed to shaping the future of XR through a child-centered and design-aware lens.

I bring to the workshop a perspective grounded in empirical research, with a focus on how game design contributes to children's safety and well-being in XR environments. Our work investigates harmful design patterns in XR games; explores how community ideation among game developers on platforms like Roblox can lead to safety risks; understands child players experiences of safety risks in such games. I aim to contribute both a designer's perspective—on how game design structure player experiences—and a child-centered perspective, informed by children's own accounts of harm in immersive games. I hope to broaden the conversation on XR ethics by emphasizing the design-level safety risks embedded in interactive and immersive XR technologies.

Through this workshop, I hope to deepen my understanding of ethical considerations in XR design and explore how existing frameworks might be adapted to better reflect children's developmental needs and lived experiences. I am especially interested in identifying design principles that can guide the creation of safer, more inclusive XR environments and in building collaborations that bridge research, practice, and policy. Ultimately, I aim to help advance a collective vision for ethical and developmentally appropriate XR design that prioritizes the safety and agency of children.

## ACKNOWLEDGMENTS

This work was partially supported by National Science Foundation under grant #2326505.